\documentclass[acmsmall]{acmart}
\AtBeginDocument{%
  }

\setcopyright{cc}
\setcctype{by}
\acmJournal{PACMHCI}
\acmYear{2026} \acmVolume{10} \acmNumber{6} \acmArticle{CSCW176}
\acmMonth{10} \acmDOI{10.1145/3817024}

\definecolor{shadecolor}{gray}{0.94}
\usepackage{framed}

\newcommand{\groupOne}{\text{Balanced condition}}
\newcommand{\groupTwo}{\text{Opposing condition}}
\newcommand{\groupThree}{\text{Reinforcing condition}}
\newcommand{\groupOneBot}{\text{Balanced chatbot}}
\newcommand{\groupTwoBot}{\text{Opposing chatbot}}
\newcommand{\groupThreeBot}{\text{Reinforcing chatbot}}

\begin{document}

\title{Understanding and Supporting Online Discussion with Opinionated Chatbots}

\author{Tianqi Song}
\email{tianqi_song@u.nus.edu}
\orcid{0000-0001-6902-5503}
\affiliation{%
  \institution{National University of Singapore}
  \country{Singapore}
}

\author{Chi-Lan Yang}
\email{chilanyang@slis.tsukuba.ac.jp}
\orcid{0000-0003-0603-2807}
\affiliation{%
  \institution{University of Tsukuba}
  \city{Ibaraki}
  \country{Japan}
}

\author{Zihan Liu}
\email{zihan.liu.lzh@gmail.com}
\orcid{0009-0003-5956-7132}
\affiliation{%
  \institution{Informatics, University of Illinois at Urbana-Champaign}
  \city{Illinois}
  \country{United States}
}

\author{Zhengtao Xu}
\orcid{0009-0003-4429-4768}
\email{xuzhengtao@u.nus.edu}
\affiliation{%
  \institution{National University of Singapore}
  \country{Singapore}
}

\author{Yibin Feng}
\email{feng.yibin@u.nus.edu}
\orcid{0009-0003-9109-8460}
\affiliation{%
  \institution{National University of Singapore}
  \country{Singapore}
}

\author{Yi-Chieh Lee}
\orcid{0000-0002-5484-6066}
\email{yclee@nus.edu.sg}
\affiliation{%
 \institution{National University of Singapore}
 \country{Singapore}}

\renewcommand{\shortauthors}{Tianqi Song et al.}

\begin{abstract}
Opinionated chatbots are increasingly present on online platforms and have the potential to shape public discourse by influencing individuals' viewpoints before they engage in discussions. Despite their growing presence, the impact of interacting with opinionated chatbots on subsequent online interactions remains largely unexplored. This study investigated how exposure to different types of opinionated chatbots, specifically those expressing opposing, reinforcing, or balanced viewpoints, affected participants' subsequent online discussions. In a controlled experiment with 83 participants, we found that interacting with an opinionated chatbot that consistently opposed participants’ arguments led to greater shifts in opinion, indicating enhanced openness to revising one’s initial stance. Conversely, participants who interacted with a chatbot that consistently reinforced their views were more likely to adopt more agreeable communication styles in subsequent conversations with others. 
Furthermore, interactions with different types of opinionated chatbots resulted in varying levels of trust, as well as different perceptions of chatbots and human interlocutors. Our findings indicate that opinionated chatbots can influence both individuals' opinions on social topics and their communication behaviors in online environments. This presents a trade-off for future designers seeking to facilitate cognitive flexibility in changing opinions while maintaining positive user experiences and trust in the chatbots during public discourse. We discuss the implications for designing opinionated chatbots to promote more constructive and less polarized online discussions.
\end{abstract}



\received{May 13, 2025}
\received[revised]{January 13, 2026}
\received[accepted]{March 17, 2026}

\maketitle

\section{Introduction}

Online discussions, facilitated by social media and other platforms, allow for widespread participation in political and social debates~\cite{shortall2022reason}. These discussions have notable benefits, such as promoting civic engagement \cite{pendry2015individual} and facilitating mutual understanding.
However, online discussion does not begin at the moment of verbal discussion among interlocutors. It is often preceded by a preparatory phase in which individuals seek background information, anticipate counterarguments, and rehearse how to articulate their positions.
This preparation is not merely an individual cognitive activity but a socially oriented one. People prepare their opinions with imagined audiences in mind, anticipating social evaluation, disagreement, and the consequences of making their views visible \cite{gayman2020varying}. As a result, information seeking and argument formulation during preparation before online discussion are inherently social processes, shaped by social expectations and anticipated reactions from others.

Although preparation has long been part of discussion practices, its role in influencing subsequent online interactions has received limited attention. As opinion formation and argument preparation are increasingly mediated by AI systems, this preparatory phase warrants closer examination.
In everyday use, search engines and AI-generated summaries often present information that implicitly or explicitly reflects particular viewpoints~\cite{sharma2024generative, spatharioti2023comparing}. 
Beyond information access, it is foreseeable that people could increasingly use AI chatbots as opinionated partners~\cite{kim2020bot, hadfi2021argumentative, tanprasert2024debate}, asking them to generate supporting or opposing arguments or to role-play alternative perspectives as part of their preparatory process.
Furthermore, prior work has shown that opinionated chatbots that consistently agree or disagree with users can help participants generate richer arguments and deepen their reasoning when preparing for interviews~\cite{nofal2025ai}, group discussions~\cite{kim2020bot, labadze2023role}, and online debates~\cite{zeng2025ronaldo, lee2024can}. 
Together, this growing adoption of \textit{opinionated AI} in argument formation highlights the emergence of AI as an active participant in the preparatory phase of online discussion.

However, \textit{how does interacting with different types of opinionated AI prior to a discussion influence participants’ subsequent interpersonal interactions?}
This question is grounded in the assumption of \textit{social influence theory}~\cite{zimbardo1991psychology}, which describes how individuals adapt their opinions, revise their beliefs, or change their behavior through social interaction. 
Importantly, social influence does not necessarily rely on relational or emotional processes. It can also operate at a cognitive level by shaping how individuals reason about an issue and evaluate arguments, in ways that resemble subtle nudges embedded within social interactions \cite{forgas2016social, bandura1977social}.
As generative AI systems increasingly exhibit socially responsive behaviors~\cite{xu2025revealing} and become harder for people to distinguish from human counterparts~\cite{reinhart2025llms}, according to social influence theory, interactions with opinionated AI may similarly exert social influence on humans.
Thus, we are curious about whether interacting with opinionated AI imposes social influence on subsequent human-human interactions.

Previous research has investigated opinionated chatbots as tools for facilitating and managing conversations, primarily by exposing users to diverse or opposing viewpoints and examining their influence on individual cognitive processes. These systems have been shown to support individual critical thinking~\cite{tanprasert2024debate, park2025assessing, lee2025conversational}, reduce confirmation bias~\cite{dingler2018biased, song2025interaction}, and address challenges such as filter bubbles and online polarization~\cite{zhang2024see, xie2025discussions}.
However, little is known about the social influence of such preparatory human–chatbot interactions on subsequent human–human communication, including how people communicate and perceive their interlocutors during interpersonal discussion.

Moving from the impact of opinionated AI at an individual level to a social one, this study developed and compared three distinct types of opinionated chatbots: one presenting arguments from both sides of the topic (balanced), another consistently opposing individuals' stances (opposing), and a third consistently reinforcing them (reinforcing). Participants interacted with one of these chatbots to prepare for a subsequent online discussion on the same topic with a human interlocutor. 
In a controlled experiment with 83 participants, we found that opinionated chatbots influenced not only participants’ stances on the topic but also their communication patterns in the following interpersonal conversations.
Specifically, while chatbots programmed to consistently oppose participants' stances tended to diminish participants' trust toward the chatbot, these interactions significantly increased the likelihood of participants altering their initial opinions in the subsequent online discussions, compared to interactions with other chatbot types. Conversely, engaging with a chatbot that consistently reinforced a participant's viewpoints led to an increased use of agreeable expressions when interacting with their human counterparts in the subsequent discussion.

The contributions of this research are twofold. First, it provides empirical evidence demonstrating that preparatory interactions with an opinionated language model influence not only users' individual perceptions and stances but also affect the outcomes of subsequent interpersonal online discussions.
Second, this work contributes to CSCW by proposing new design perspectives on leveraging opinionated chatbots as pre-discussion facilitators, highlighting their potential to enable constructive online deliberation through subtle, implicit forms of influence.
\section{Related Work}

\subsection{Supporting Online Discussions with Digital Solutions}
\label{sec:motivation}
People exchange opinions or engage in deliberations online.  
However, empirical evidence has consistently shown that online environments do not always foster rational debate and deliberation, and can even fuel misunderstanding and conflict~\cite{argyle2023leveraging}. This shortfall can be attributed to several factors: users often gather in echo chambers, seeking out information that reaffirms their existing views, which hinders a genuine understanding and appreciation of opposing viewpoints~\cite{mutz2006hearing}. Furthermore, user-generated information on the Internet is often emotionally charged, particularly in discussions about politics or values. This can escalate tension and polarization by provoking defensive rather than reflective responses~\cite{kwon2017swearing}.
Compounding these issues are intrinsic features of online communication, such as anonymity and the absence of direct feedback, which can lead to toxic disinhibition, manifesting in impulsive and uncivil behavior~\cite{suler2004online}. 
These challenges highlight the complexity of having meaningful and inclusive dialogues in the online public discourse.

To address the challenges of scaling high-quality online discussions, researchers have explored various technological solutions \cite{shortall2022reason}, primarily focusing on two approaches: argumentation tools and facilitation methods. 
Argumentation tools are designed to enhance the structure and quality of arguments within online discussions by providing automated feedback, as demonstrated in tools like Arguetutor \cite{wambsganss2021arguetutor} and LegalWriter \cite{weber2024legalwriter}, which significantly improve students' argumentative contributions \cite{wambsganss2022enhancing}. 
Facilitation methods, on the other hand, aim to guide and moderate interactions to improve discussion dynamics, such as Lee et al.'s SolutionChat system, which visualizes discussion stages and supports moderators with context-aware messages \cite{lee2020solutionchat}.

Among these technological interventions, AI chatbots have emerged as promising tools for supporting online discussions~\cite{mohammed2019towards}, particularly in facilitating structured dialogue~\cite{lee2020solutionchat}, surfacing supporting evidence for arguments~\cite{zeng2025ronaldo}, moderating tone~\cite{kim2021moderator, zeng2025ronaldo}, and increasing equal participation~\cite{hadfi2023conversational}.
Prior studies have explored chatbots as moderators or assistants that scaffold online deliberation. For instance, “DebateBot” was designed to support quieter voices in discussions by guiding argumentation~\cite{kim2021moderator}, while other interventions focused on reducing political tension by offering real-time, evidence-based prompts~\cite{argyle2023leveraging}. Similarly, chatbots presenting counter-attitudinal content have been found to reduce selective exposure and increase openness to opposing views~\cite{zarouali2021overcoming}. These systems effectively improve discussion quality and inclusiveness by creating environments that are more welcoming to ideological diversity, thereby advancing the broader goal of fostering deliberative democracy in online public spheres.

\subsection{Opinionated AI Chatbots as Preparatory Tools}
\label{sec:chatbot-discussion}
The use of opinionated AI chatbots in online discussion is growing, with people increasingly engaging these systems to role-play different positions as a way of preparing for real discussions in everyday practice.
In research settings, LLM-based chatbots have likewise been used as preparatory tools for online discussion. A common practice is to employ chatbots such as ChatGPT to generate arguments that support a given statement, with the chatbot consistently taking an agreeing stance to help users articulate and elaborate their views~\cite{zeng2025ronaldo, hadfi2023conversational}. In contrast, another growing body of work positions chatbots as debate partners or devil’s-advocate agents that present counterarguments to stimulate critical thinking~\cite{tanprasert2024debate, lee2024can, park2025assessing, lee2025conversational}. These systems have been studied across diverse contexts, including online discussions~\cite{tanprasert2024debate}, educational debate practice~\cite{lee2024can}, and group decision-making~\cite{lee2025conversational}.

In such cases, opinionated AI chatbots no longer function solely as tools for argument generation but also exert influence through their social and interactional roles.
For example, \citeauthor{tanprasert2024debate} found that debate chatbots shaped both users’ critical thinking and their perceptions of the chatbot, with communicative styles producing distinct effects: persuasive rhetoric encouraged deeper self-reflection, while eristic rhetoric led users to reinforce their own counterarguments~\cite{tanprasert2024debate}.
Similarly, devil’s-advocate systems powered by LLMs have been found to amplify minority voices and enhance psychological safety in group decision-making~\cite{lee2025conversational, lee2025amplifying}.
Yet, this body of work has primarily examined the immediate dynamics of human–AI interaction and outcomes such as individual critical thinking or group inclusivity. 
Much less is known about how prior encounters with opinionated chatbots might influence subsequent human-to-human discussions, or how such interactions may subtly influence users' own communication patterns.

This gap is particularly significant in online discussions, as the way communication occurs can be as important as the opinions being expressed.
Prior CSCW research highlights that good faith disagreement and effective conflict management are essential for deliberative democracy and for sustaining constructive online relationships~\cite{baughan2021someone}. Building on this perspective, an important but underexplored question is whether chatbots that consistently use different communicative styles can influence users’ interpersonal behaviors in ways that go beyond opinion change. While previous studies demonstrate that interacting with opinionated language models can subtly shift users’ beliefs when people are expressing opinions individually~\cite{jakesch2023co}, it remains unclear whether interacting with language models that use different communicative styles also influences how people express agreements or disagreements in subsequent human conversations.
This research gap motivates our investigation: \textit{What if AI chatbots, designed to represent different perspectives, consistently express opposing, reinforcing, or balanced viewpoints with people to prepare them for viewpoint exchanges? How might this influence the participants in engaging in online discussion?}

\subsection{Cognitive Foundations of Social Influence}
Social influence refers to the ways individuals’ attitudes, behaviors, and perceptions are shaped through interaction with others~\cite{cialdini2004social}. Although it is often associated with emotionally or interpersonally salient contexts, social influence can also operate through information-processing contexts that do not rely on social pressure, affiliation, or affective bonding~\cite{forgas2016social}. In such cases, influence arises as individuals engage with socially sourced information, e.g., making sense of others’ perspectives, arguments, and reasoning~\cite{deutsch1955study}, offering a useful lens for understanding online discussion preparation.

While social influence captures the outcome-level phenomenon of how individuals change in social contexts, the mechanisms underlying such influence can vary~\cite{forgas2016social}. One important mechanism operates through \textbf{cognitive processes} that shape how individuals attend to information, evaluate arguments, and interpret social cues~\cite{bandura1977social}. For example, prior work shows that repeated exposure to others’ ways of framing and evaluating situations can lead individuals to internalize salient cues and judgment criteria, with effects that persist beyond the original interaction context~\cite{bryant2009media, bandura1963imitation}.

As AI systems become increasingly prevalent and socially embedded, researchers have begun to examine how social influence extends beyond interpersonal interaction to human–AI interaction~\cite{prinz2022smiling, riva2022social, li2025we}. 
For example, \citeauthor{jakesch2023co} investigated social influence in the context of an AI writing assistant~\cite{jakesch2023co}, finding that users who received stance-specific suggestions from an opinionated AI during collaborative writing were more likely to adopt those stances themselves, demonstrating a nudging effect on user behavior. 
Recently, \citeauthor{song2025multi} examined social influence from multiple AI chatbots and found that users tended to conform when these agents presented similar arguments~\cite{song2025multi}.

However, existing work has primarily focused on immediate or localized effects of human–AI interaction, such as opinion change during the interaction itself~\cite{jakesch2023co, song2025multi}.
Much less is known about whether cognitively oriented influence introduced during preparatory interaction with AI carries over to shape subsequent human–human communication. In particular, it remains unclear whether interacting with opinionated chatbots influences how users attend to, evaluate, and perceive their interlocutors during later interpersonal discussions. Addressing this gap is critical for understanding the broader social consequences of using opinionated AI systems as preparatory tools for online discussion.

\subsection{Research Questions}
This study seeks to understand how people perceive different opinionated chatbots when having online discussions and the impact of these human-bot interactions on subsequent interpersonal discussions. 
We focus on three types of opinionated chatbots: \emph{reinforcing}, \emph{opposing}, and \emph{balanced}. Reinforcing and opposing chatbots represent two dominant ways people use opinionated AI in practice, i.e., seeking affirmation of existing views or seeking challenge through counterarguments, while balanced chatbots provide information without adopting a directional stance and serve as a baseline condition. 
Contrasting these three conditions allows us to examine how different forms of opinionated exposure during preparation influence later interaction, rather than evaluating isolated chatbot behaviors.

Motivated by social influence theory and prior work on preparatory practices in discussion, we adopt an exploratory approach to examine how influence may manifest across multiple dimensions. 
Specifically, we investigate not only opinion change, but also communication patterns and user perceptions, as these factors jointly shape the quality and dynamics of online interpersonal discussion.

\subsubsection{Impact of Opinionated Chatbot on Opinion Change (RQ1)}
We began by addressing a fundamental yet important question: to what extent can different opinionated chatbots influence users’ opinions in the context of online discussion? 
This question is theoretically grounded in informational social influence, which suggests that exposure to persuasive information can lead individuals to revise their beliefs. Accordingly, we hypothesize that chatbots expressing opposing or reinforcing viewpoints may shift users’ opinions, whereas exposure to balanced viewpoints may lead to little or no opinion change.
By examining how users respond to chatbots that consistently agree or disagree with their stance, we aim to understand whether such AI-mediated opinion exposure can promote or reduce openness to alternative perspectives in subsequent human-to-human interactions.

Prior research has shown that language models can subtly influence users' opinions in individual settings, as measured through self-reported surveys~\cite{jakesch2023co}. However, it remains unclear whether and how these effects carry over into subsequent interpersonal interactions, such as online discussions with other humans.
Moreover, the literature offers conflicting evidence on how exposure to opposing views affects opinion dynamics. Some studies suggest that disagreement can foster critical thinking, reflection, and even consensus-building~\cite{tanprasert2024debate}, while others argue it can reinforce existing beliefs and exacerbate polarization through biased information processing~\cite{lord1979biased, knobloch2009looking}. This tension highlights the need for additional research to explain whether and why different opinionated chatbots may shift users' opinions.
Therefore, we asked our first research question:

\textbf{RQ1}: How does interacting with reinforcing, opposing, or balanced chatbots during preparation influence users’ opinions during subsequent interpersonal discussion?

\subsubsection{Impact of Opinionated Chatbot on Communication Pattern (RQ2)}
Beyond the focus of RQ1, our research also aims to examine the impact of such interactions on subsequent interpersonal communication. 
This inquiry was grounded in previous studies in human–human communication, which shows that individuals often adjust their speech and nonverbal behavior to align with their conversational partners~\cite{giles1991accommodation, giles2023communication}. In the context of interacting with technology, recent studies have revealed the conceptual alignment in human-robot interaction~\cite{cirillo2022conceptual} and lexical alignment with computers~\cite{shen2023effects}.
These insights suggest a high likelihood of language convergence after discussing with chatbots designed to present specific stances—agreeing, disagreeing, or neutral- thus priming users to approach subsequent interpersonal dialogues with a more open and reflective attitude, potentially enhancing the quality of discourse.
Consequently, we asked our second research question:

\textbf{RQ2}: How does prior interaction with different types of opinionated chatbots shape users’ communication patterns when engaging in subsequent interpersonal discussion?

\subsubsection{Impact of Opinionated Chatbot on User Perceptions (RQ3)}
Lastly, we investigated whether and how interacting with opinionated chatbots might shape users’ perceptions of human interlocutors in subsequent interpersonal conversations, and how these impressions may relate to the ones formed during prior human–chatbot interactions~\cite{hatfield1993emotional}. This topic remains largely underexplored in the context of human–AI interaction, particularly regarding the subsequent social interactions after chatbot-mediated preparation for interpersonal exchange. Given the exploratory nature of this question, we employed both surveys~\cite{balietti2021reducing} and open-ended responses to gain richer insight into users’ subjective impressions.

Thus, we asked:

\textbf{RQ3}: How do users perceive and interpret different opinionated chatbots, and how do these perceptions relate to their subsequent interpersonal interaction experiences?

\section{Method}

\begin{figure}
    \centering
    \includegraphics[width=0.9\textwidth]{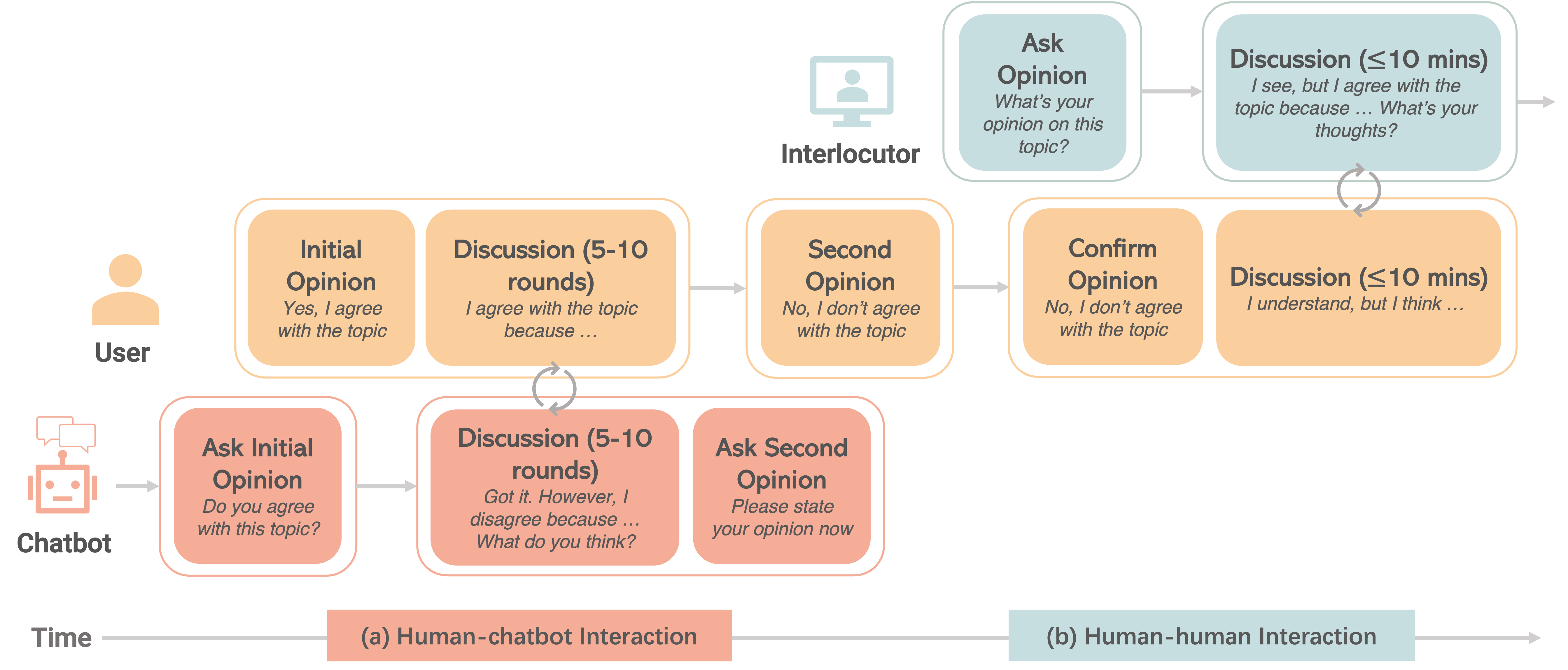}
    \caption{The conversation design comprises two parts: (a) the participant's interaction with the chatbot and (b) the participant's interaction with an interlocutor. The first part begins with the chatbot asking for the participant's initial opinion and ends when the participant inputs "stop." The second part starts with the interlocutor asking for the participant's opinion on the same topic and concludes when the participant and interlocutor reach a consensus.}
    \label{fig:chatbot-design}
\end{figure}

\subsection{Experiment Design}

To answer our research questions, we designed and compared three conditions with a between-subject design: interacting with a \groupOneBot{}, \groupTwoBot{}, and \groupThreeBot{}. Participants were asked to discuss a social issue with one of three types of chatbots depending on the assigned condition. Then they proceeded with the discussion with another interlocutor in text chat. In \groupOne{}, participants interacted with a chatbot that maintained a balanced stance, neither opposing nor endorsing any viewpoints presented by the participants. In \groupTwo{}, the chatbot consistently opposed the views expressed by the participants. In \groupThree{}, the chatbot consistently supported the viewpoints expressed by the participants.

We selected a real-world topic, \textit{``whether economic growth can help environmental protection''}, for participants to discuss in the experiment. This topic was chosen because it is a widely debated social issue, encouraging participants to share diverse perspectives. Furthermore, the topic does not have a definitive right or wrong answer \cite{vcabelkova2023environmental}, making it particularly effective for eliciting varied attitudes based on individuals' beliefs. This characteristic also reduces the impact of potential factuality hallucinations by LLMs \cite{huang2023survey}, as the discussion relies on subjective reasoning rather than factual accuracy, making it an appropriate choice for a controlled experimental setting.

\subsection{Chatbot design}
We used a combination of script-based instructions and the GPT-3.5 API to implement \textit{Balanced condition, \groupTwo{}}, and \textit{\groupThree{}}. The conversation flow is depicted in Figure \ref{fig:chatbot-design}. The script-based instructions were used to guide the procedure, while the GPT-3.5 API was used for dynamic and engaging interactions and discussions.

The discussion begins with the chatbot prompting participants to express their opinions and corresponding arguments. During each round of the discussion, the chatbots were prompted to demonstrate an understanding of the users' arguments and provide a reaction based on the corresponding conditions:

\begin{itemize}
    \item \groupOne{}: The chatbot would propose two arguments, one supporting and one opposing the users' arguments, such as \textit{"I understand what you mean. One argument in support of your view is... On the other hand, one argument against your view is..."}
    \item \groupTwo{}: The chatbot would express disagreement with the user's opinions, stating \textit{"I understand, but I disagree, because..."} and propose an argument to support their own viewpoint. 
    \item \groupThree{}: The chatbot would express agreement with the user's opinions, stating \textit{"I completely agree with you, because..."} and propose an argument to further support their shared viewpoint. 
\end{itemize}

The prompt design followed the guidelines outlined in the OpenAI Prompt Engineering Document \footnote{https://platform.openai.com/docs/guides/prompt-engineering}, defining the chatbot's role, tone, and actions. To maintain consistency in the chatbot's responses, we set the API's temperature parameter to 0.2, ensuring stable and predictable interactions. Prior to the experiment, we conducted multiple pilot studies to verify the prompt's effectiveness and refine its functionality. During the experiment, we also monitored response quality to maintain alignment with the intended chatbot behavior.

Note that, the prompt design for both the opposing and reinforcing chatbots enabled them to appropriately respond to neutral or mixed positions by either affirming or challenging the participant’s arguments directly. 
Detailed parameters and prompt configurations are provided in Appendix \ref{app:chatbot}.

\begin{figure}
    \centering
    \includegraphics[width=\linewidth]{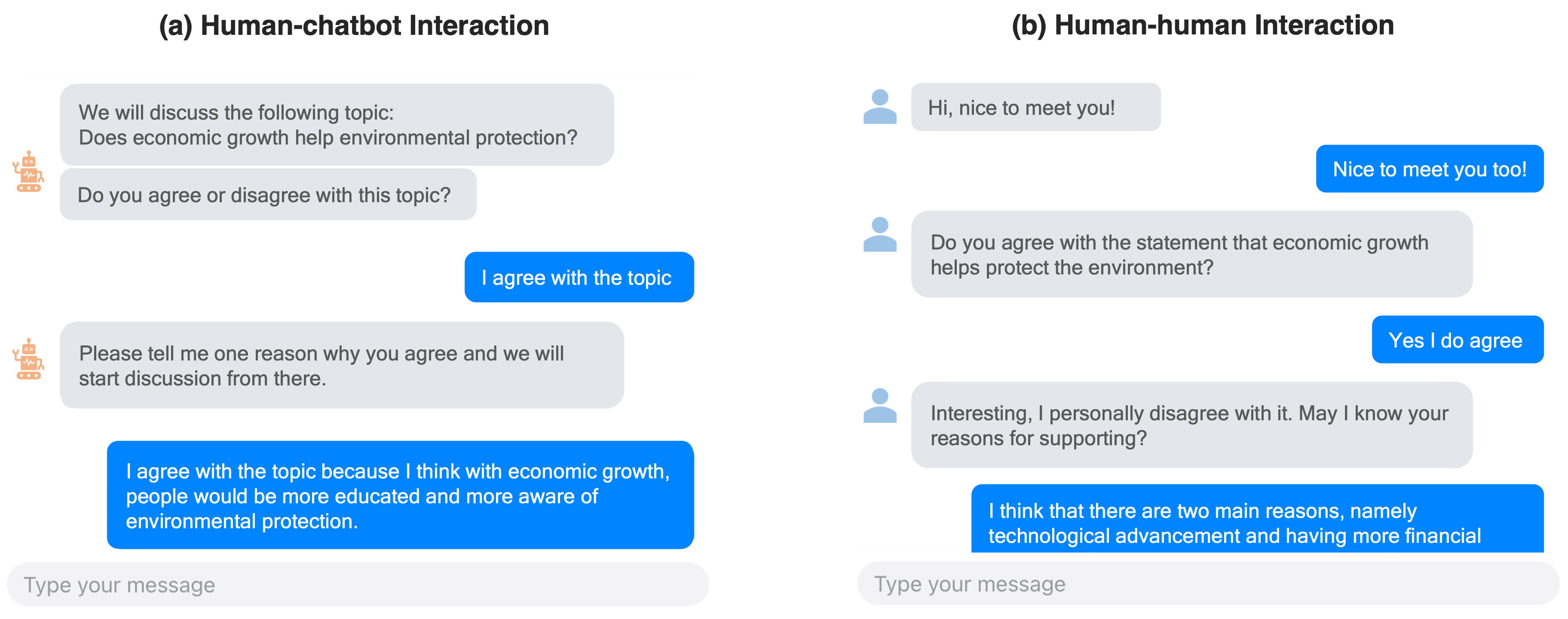}
    \caption{Illustrative example of the chatting interface from the participants' perspectives: (a) interaction with a chatbot and (b) interaction with a human interlocutor. The same user interface design was utilized for both scenarios to ensure consistency.}
    \label{fig:screenshot}
\end{figure}

\subsection{Participants}
We recruited participants through the online platform Telegram, with eligibility criteria set for students in specific institutions aged 21 and above.
In the end, we collected a total of 90 responses, and 7 were removed due to a misunderstanding of the task or inconsistent answers given during the task. We ended up with 83 valid responses, with 29 participants (F:15, M:14) in \groupOne{}, 27 participants (F:14, M:13) in \groupTwo{}, 27 participants (F:14, M:13) in \groupThree{}.
The average ages for each condition were as follows: \groupOne{} = 26.33 (SD = 5.62), \groupTwo{} = 25.93 (SD = 4.91), and \groupThree{} = 25.40 (SD = 5.46).
The educational background varied among the participants: 37 possessed a bachelor's degree, 21 had a high school diploma or equivalent, 15 had some college experience without a degree and 10 had a master’s degree or higher.

The entire experiment duration was approximately one hour, with participants receiving a reimbursement of 12 US dollars per hour for their involvement.

\subsection{Procedure}
\begin{figure}
    \centering
    \includegraphics[width=\textwidth]{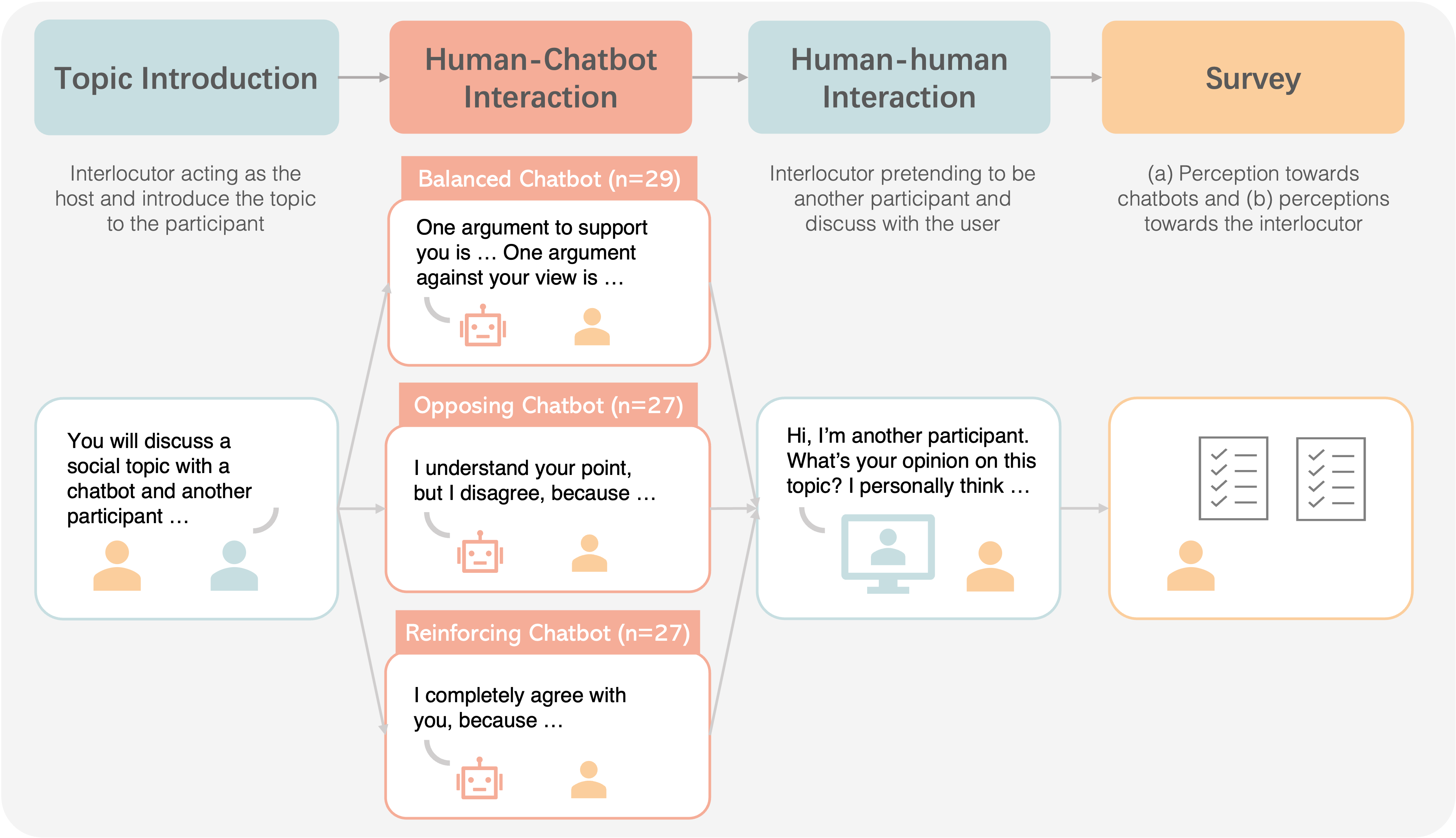}
    \caption{The study procedure includes four parts: (1) Topic Introduction. A research assistant introduces the study procedure and social topic to the participant online; (2) Human-Chatbot Interaction. Participants interact with one of three chatbots for 5-10 rounds of conversation on the topic; (3) Human-Human Interaction. Participants interact with an interlocutor, played by a research assistant, who initially disagrees with the participant using script-based arguments and then gradually reaches a consensus; (4) Survey. Participants complete a survey covering two aspects: (a) their perceptions of the chatbots and (b) their perceptions of the interlocutor.}
    \label{fig:study-design}
\end{figure}

As shown in figure \ref{fig:study-design}, the experiment was structured as follows: When signing the consent form, participants received online instructions outlining the entire procedure. They were informed that they would first engage in a discussion with an AI-powered chatbot, followed by a discussion with a human participant on the same topic.

Each participant joined the study with a confederate throughout the experiment. The confederate had two primary roles: 1) guiding participants through the study process as the host and 2) acting as the interlocutor during the online discussion. 
Note that the entire interaction was text-based, with no video, audio, or facial expressions, so participants were unaware that the host and the discussion partner were the same person.
Participants began the experiment by joining a Zoom meeting with the confederate, sharing their screens to allow observation of their actions and to ensure timely identification and resolution of any unexpected issues.

After familiarizing themselves with the process, participants first engaged in conversations with a chatbot for 5-min. During this stage, the chatbot introduced the topic, provided background information, and asked participants to express their stances on the topic.
Once the participants had chosen their stance, the chatbot prompted them to provide their initial arguments in support of their position. The chatbot responded to participants with at least one supporting argument using different attitudes depending on the assigned conditions. Participants were encouraged to engage in multiple back-and-forth conversations with the chatbot, and the discussion would continue for a minimum of five rounds and a maximum of ten rounds.

Following the interaction with the chatbot, participants engaged in an online discussion on the same topic with the confederate. Participants were explicitly informed about this transition, with the online website displaying the message: \textit{
“Now you are going to discuss with a real person.”}

During the conversation, the confederate was instructed to consistently adopt a disagreeing stance toward the participant for approximately 10 minutes, while gradually working toward consensus. Unlike the initial chatbot interaction, the discussion in this stage concluded only when both parties reached a shared agreement.

To minimize variability and potential bias across sessions, the confederate followed a predefined interaction script. 
The script specified a standardized set of arguments both supporting and opposing the topic, ensuring consistency across participants. 
During the discussion, the confederate was instructed to consistently disagree with the participant using the provided arguments and to gradually move toward consensus as the discussion progressed. The confederate was instructed to rely exclusively on the provided arguments, which could be copied and pasted with minor edits. 
Importantly, the confederate was not informed of the study hypotheses or experimental conditions beyond the required stance-taking behavior, and did not have access to participants’ prior interactions with the chatbot.
Detailed scripts and instructions provided to the confederate are included in Appendix~\ref{app:scripts}.

Afterward, participants would be asked to complete the online survey, sharing their experiences regarding the discussion topic for about 10 minutes.

The experiments were conducted using an online platform with a chatting interface integrated through the Uchat Chat Widget \footnote{https://docs.uchat.com.au/flow-builder/tools/widgets.html}. This widget enabled the incorporation of a flexible chat UI into the website. Participants interacted with either a chatbot or a human interlocutor through the same interface, as shown in Figure \ref{fig:screenshot}. Different avatars were used to distinguish between the two identities.

\subsection{Measurements}

\subsubsection{Opinion Shift Analysis}
\label{method:opinion-change}
We asked participants to express their opinions before and after interacting with the chatbot. 
To determine participants' stances, they were asked to respond to the chatbot's question,
\textit{"Do you agree or disagree with this topic?"}
While participants could simply answer “yes” or “no,” they were also encouraged to elaborate on their reasoning. This open-ended format was intended to capture not only participants’ stances but also the arguments and thought processes underlying those stances.

To analyze participants’ responses, we categorized their selected statements into three groups: "supporting the topic", "opposing the topic", or "no clear stance", allowing for systematic comparison.
To ensure the reliability of this qualitative coding, two independent coders annotated 27 randomly selected responses (30\% of the total 83), yielding a strong inter-rater reliability score ($\kappa = .87$). Disagreements occurred only in a few cases and were primarily related to determining whether a response lacked a clear stance.

A difference in their chosen statement before and after interacting with the chatbot was considered an opinion shift. 
For example, if a participant chose \textit{"I agree with the topic"} at first, and changed to \textit{"I disagree with the topic"} or \textit{"I am not sure"} after interacting with the chatbot, it would be labeled as "opinion shift".
The proportion of opinion shifts was calculated by dividing the number of participants who changed their opinion by the total number of participants in that condition. 
For example, if there are $n$ participants in \groupOne{} and $x$ \groupOne{} participants changed their opinion, the proportion of opinion shifts in \groupOne{} was calculated as $x/n$.

\subsubsection{Communication Pattern Analysis}
\label{method:conversational-log}
The unit of analysis was a round of conversation, where each round starts with the interlocutor's arguments and ends with the participant's arguments. For instance, one round of conversation from P25 (\groupOne{}, M) begins with the interlocutor saying \textit{"Actually, most of the industries that cause pollution have moved their factories to other countries, like Vietnam. So the damage is not reduced"}, to which the user responds \textit{"but I feel that this might not be practical because the government cannot just randomly give away large amounts of money for this"}.
In total, there were 810 conversation rounds from 83 participants in the second part of the experiment (\autoref{fig:study-design}, Part 2), with an average of 9.76 rounds of conversation between the participants and their interlocutors.

To examine how participants' language usage with their interlocutors was influenced by their prior interaction with the chatbots, we labeled the participants' conversations with their interlocutors according to their communication patterns exhibited. 
We first identified conversation rounds that were small talk (i.e., \textit{"Hi, my name is xxx"}), irrelevant to the discussion topics (i.e., \textit{"Maybe we can stop here"}), or did not contain valid arguments (i.e., \textit{"Do you know Paris agreement?"}). These were labeled as "NA".
Following the removal of "NA" conversations, a total of 502 rounds of conversations were identified (\groupOne{}: 170, \groupTwo{}: 160, \groupThree{}: 172). The average number of rounds of conversations labeled for communication patterns is 6.05 per participant.

We categorized participants' communication patterns when expressing their attitudes to their interlocutors:
\begin{itemize}
    \item Showing agreement: The participant explicitly agrees with the interlocutor (e.g., \textit{"I agree"}, \textit{"You make a good point"}).

    \item Showing disagreement: The participant clearly expresses disagreement (e.g., \textit{"I disagree"} or \textit{"I do not feel that there"}, or simply disagreed by saying \textit{"However ..."}).

    \item Unclear stance: The participant’s response is ambiguous or mixed, without a clear indication of agreement or disagreement (e.g., \textit{"I get where you are coming from"}, \textit{"That may be true"}, and \textit{"Yes and no"}).
\end{itemize}

Regarding the communication pattern, two coders coded the same 150 rounds of conversation, representing 30\% of the total dataset ($\kappa = .82$). 
They then deliberated on addressing any differences in coding and reaching a consensus. This process allowed them to refine their approach.

\subsubsection{Open-Ended Comments.}
\label{method:open-ended comments}
In addition to the survey and conversation log, we also included open-ended questions to capture users' thoughts about their interactions with the chatbot and interlocutor. Specifically, we asked participants questions about 1) their impression of the chatbots, and 2) their impression of the interlocutors.

We analyzed these responses using an inductive thematic analysis approach. The unit of analysis was each participant’s complete response to an open-ended question. Two researchers first independently familiarized themselves with the data by reading all responses multiple times. They then jointly reviewed and iteratively coded an initial subset of 25 responses to develop a preliminary coding scheme.
During this phase, codes were generated at a semantic level to capture participants’ explicit evaluations and descriptions. For example, in response to the question \textit{``What is your impression of the chatbot, and why?''}, initial codes included functional attributes (e.g., \textit{``useful,'' ``informative,'' ``logical reasoning''}) and social characteristics (e.g., \textit{``kind,'' ``friendly,'' ``assertive''}).
After establishing the initial codebook, the two researchers independently applied the coding scheme to the remaining responses. They then met regularly to compare coding results, resolve discrepancies through discussion, and refine code definitions when necessary. This iterative process continued until the coding scheme stabilized and satisfactory inter-rater reliability was achieved ($\kappa = .89$).
The resulting themes were used to contextualize and interpret the quantitative findings by illustrating how participants made sense of the chatbots’ stances and how these perceptions carried over into subsequent human–human discussions.

\subsubsection{Survey}

To examine how preparatory interactions with opinionated chatbots shape participants’ experiences and subsequently influence interpersonal discussion, we employed survey measures adapted from prior work on communication quality and trust in mediated interaction~\cite{liu2010quality, harris2011perceived}. 
These measures were selected to capture two theoretically relevant dimensions through which social influence may manifest in online discussion contexts: users’ \emph{perceptions of conversational partners} and their \emph{evaluation of interaction quality}.

Specifically, we focused on perceived trust and communication quality because prior research has shown that both constructs play a central role in shaping how individuals engage in discussion, evaluate information, and respond to disagreement. Trust in conversational partners is closely linked to perceived credibility, openness to influence, and willingness to engage in deliberation~\cite{walther2005rules}. 
Communication quality, in turn, has been associated with deeper information processing, constructive engagement, and openness to opposing viewpoints in online discussion settings~\cite{diakopoulos2011towards, himmelroos2014deliberation}. 
Together, these dimensions allow us to examine not only whether opinionated chatbots influence users’ attitudes, but also how they shape users’ perceptions of interaction and discussion partners, which is central to our research questions.

Participants rated all items on a seven-point Likert scale (1 = \textit{"Totally disagree"}, 7 = \textit{"Totally agree"}). For each construct, we computed the average score across all items and used these composite measures in subsequent analyses.

For statistical analysis, we calculated the average score across all items within each construct.

\textbf{Trust.}
The measurement of perceived trust in the \textit{chatbot} was based on the dimensions of access to quality information and perceived impartiality, as identified in previous research \cite{harris2011perceived}. Some example questions are \textit{"The advice given by the chatbot seemed credible"} and \textit{"The advice given by the chatbot seemed objective (i.e. no hidden agenda)"}.

\textbf{Communication Quality.}
The quality of communication with the \textit{chatbot and interlocutors} was taken from previous research \cite{liu2010quality} and we rephrased the questions a bit to fit our specific scenarios. 
Examples of questions for chatbot communication quality are \textit{"The overall communication quality of the chatbot is good"} and examples of questions for interlocutors are \textit{"The overall content quality of the real person is good"}.
\section{Results}

The results are organized into three parts, which are summarized in Table~\ref{tab:result-summary}.

\begin{table}
    \centering
    \small
    \renewcommand{\arraystretch}{1.3} 
    \begin{tabular}{p{0.25\textwidth}p{0.65\textwidth}}
    \toprule 
      \textbf{Research Question} & \textbf{Findings} \\ 
      \midrule
        \textbf{RQ1}: How does interacting with opinionated chatbots influence users’ opinions on a topic? & 
        \textbf{Opinion Shift} \newline
        - All three types of opinionated chatbots led to some level of opinion change (Balanced: 31\%, Opposing: 59\%, Reinforcing: 33\%). \newline
        - Participants who interacted with opposing chatbots exhibited significantly more opinion change compared to those who interacted with reinforcing or balanced chatbots. (Section \ref{sec:opinion-shift})\\ \hline
    
      \textbf{RQ2}: How does interacting with opinionated chatbots influence users’ subsequent communication patterns? & 
      
      \textbf{Communication Patterns} \newline
      - Participants who interacted with reinforcing chatbots were significantly more likely to use agreeable expressions during the follow-up human-human conversation (Reinforcing: 49\%), compared to those in other conditions (Balance: 39\%, Opposing: 31\%). \newline
      - The use of disagreeable expressions did not differ significantly across the three conditions (Balanced: 11\%, Opposing: 14\%, Reinforcing: 12\%). (Section \ref{sec:communication-pattern})
      \\ \hline
      
        \textbf{RQ3.a}: How does interacting with opinionated chatbots influence people’s perceptions towards the chatbots? & 

        - Participants reported higher levels of trust with balanced and reinforcing chatbots, while opposing chatbots were associated with significantly lower trusts. (Section \ref{sec:user-perception-survey}) \newline
        - Participants attributed different qualities to the chatbots. Opposing chatbots were frequently described as demonstrating stronger reasoning, which prompted users to reconsider their own views and, in some cases, change their minds. In contrast, Reinforcing chatbots were more often characterized by social traits, such as being friendly and kind. (Section \ref{sec:chatbot-impression-qualitative}) \newline
        \\ 
        \textbf{RQ3.b}: How does interacting with opinionated chatbots influence users’ perceptions of the following human interlocutor? 
        & 
        - Participants who interacted with different chatbot types became attuned to specific characteristics of the human interlocutor. Participants who interacted with the Opposing chatbot tended to focus more on the quality of the interlocutor’s reasoning and argument. By contrast, those who interacted with the Reinforcing chatbot used more descriptions relating to friendliness and interpersonal warmth. (Section \ref{sec:interlocutor-impression-qualitative}) \\ \bottomrule
    \end{tabular}
    \caption{Summary of findings across research questions}
    \label{tab:result-summary}
\end{table}

\subsection{The Influence of Opinionated Chatbots on Participants' Opinion Shift (RQ1)}

\label{sec:opinion-shift}
\begin{figure}
    \centering
    \includegraphics[width=0.7\textwidth]{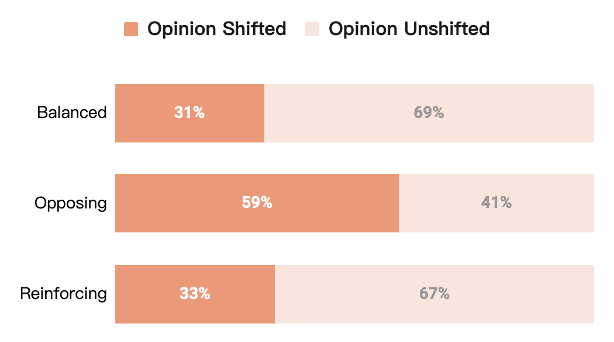}
    \caption{\%(Opinion Shift) After Discussing with Chatbots. Bar chart illustrating the ratio of participants whose opinions changed in each condition. The bar width indicates the proportion of participants who shifted or did not shift their opinions. The ratio is determined by dividing the number of participants who changed their opinions by the total number of participants in each condition. The conditions are ordered from top to bottom as \groupOne{}, \groupTwo{}, and \groupThree{}.}
    \label{fig:opinion-shift}
\end{figure}

As introduced in section \ref{method:opinion-change}, we conducted an analysis of the chatbot conversation logs of three different groups to examine the occurrence of opinion shifts among participants. 
Among the 83 participants, most participants initially expressed either support (n=42, 51\%) or opposition (n=37, 45\%) to the topic, with only a few taking a neutral stance (n=4, 5\%). The proportion of participants who changed their opinion after interacting with the chatbot was 31\% in \groupOne{} (n=9), 59\% in \groupTwo{} (n=16), and 33\% in \groupThree{} (n=9), as illustrated in Figure~\ref{fig:opinion-shift}.

Chi-square analyzes showed that participants in \groupTwo{} exhibited a higher tendency to change their opinions after discussions with chatbots than in \groupOne{} ($\chi^2 =4.51$, p<.05). There was no significant difference in opinion change between \groupOne{} and \groupThree{}, as well as between \groupTwo{} and \groupThree{} based on the results of the Chi-square analysis.
This finding suggests that \textbf{participants shifted more from their original opinions} by interacting with chatbots that always disagreed.

\subsection{The Influence of Opinionated Chatbots on Participants' Communication Patterns (RQ2)}
\label{sec:communication-pattern}

\begin{figure}
    \centering
    \includegraphics[width=0.7\textwidth]{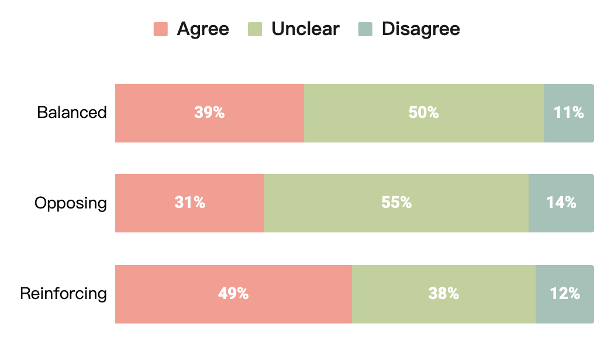}
    \caption{{\%(Communication Patterns) Between Participants and Interlocutors. Bar chart showing the ratio of conversation rounds where participants used "agree," "unclear," or "disagree" expressions. The bar width represents the proportion of conversation rounds employing these expressions. The ratio is calculated by dividing the number of conversation rounds using each expression by the total number of conversation rounds in each condition. The conditions are listed from top to bottom as \groupOne{}, \groupTwo{}, and \groupThree{}.
    }}
    \label{fig: agree level}
\end{figure}

As introduced in section \ref{method:conversational-log}, we examined the influence of chatbots' intervention on subsequent human-human communication by manually coding conversation logs between participants and interlocutors.
Figure \ref{fig: agree level} provides a summary of the results, where the majority of discussions are characterized by agree and neutral expressions, with only a small proportion of disagree.
We can notice a clear trend towards \textbf{increased agreement expressions in conversations} after the interaction with chatbots that always agreed, and a decrease in agreement in conversations following interactions with \groupTwoBot{}.

To statistically assess whether communication patterns differed across conditions, we conducted a chi-square test of independence on a 3 (group) × 3 (communication pattern) contingency table. The omnibus test revealed a significant association between chatbot condition and subsequent communication patterns ($\chi^2(4)=12.64$, p=.013), indicating that the distribution of agreement, neutrality, and disagreement varied across groups.

To further examine which group differences contributed to this association, we conducted post-hoc pairwise chi-square tests between conditions with Bonferroni correction ($\alpha$ = .0167). These analyses showed that \textbf{the distribution of communication patterns differed significantly between the \groupTwo{} and \groupThree{} conditions} ($\chi^2(2)=11.82$, Bonferroni-adjusted p=.008). In contrast, differences between the Balanced and Opposing conditions, as well as between the Balanced and Reinforcing conditions, were not significant after correction.

Descriptively, discussions following interactions with the Reinforcing chatbot exhibited a higher proportion of agreement expressions (49\%) than those following interactions with the Opposing chatbot (31\%), whereas the frequency of disagreement remained relatively low across all conditions. These results suggest that exposure to consistently reinforcing versus opposing chatbot stances can differentially shape the distribution of agreement and neutrality in subsequent human–human discussions, while explicit disagreement remains comparatively infrequent.

\subsection{The Influence of Opinionated Chatbots on Participants' Perceptions (RQ3)}

\subsubsection{Trust and Communication Quality}

\begin{figure}
    \centering
    \includegraphics[width=\textwidth]{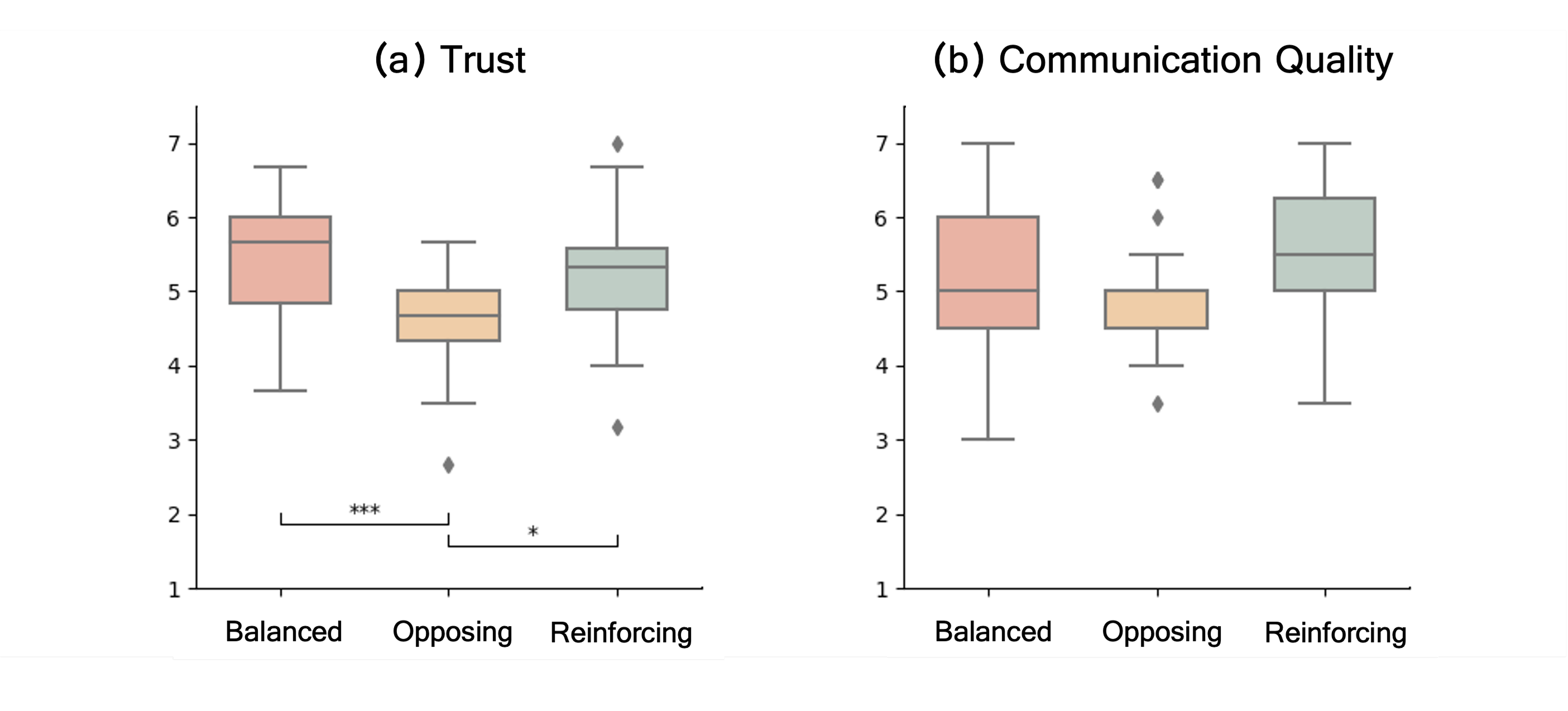}
    \caption{Perceived Trust (left) and Communication Quality (right). Box plots showing the measured values of the participants’ perceived trust towards chatbots and the perceived quality of communication with chatbots. In each boxplot, from left to right, are \groupOne, \groupTwo and \groupThree. \ensuremath{*}p \ensuremath{\leq} 0.05; \ensuremath{**}p \ensuremath{\leq} 0.01; \ensuremath{***}p \ensuremath{\leq} 0.001}
    \label{fig:trust}
\end{figure}

\label{sec:user-perception-survey}

We employed one-way ANOVA to compare participants' perceived trust towards chatbots and the quality of communication with them across three conditions. 
To account for multiple omnibus tests across outcome variables, we applied a Bonferroni correction ($\alpha$ = .0125).
After correction, the analysis revealed a significant group effect on perceived trust (F(2, 80) = 10.54, p < .001, Fig. \ref{fig:trust}a), whereas the effect on communication quality did not remain significant after correction (F(2, 80) = 3.71, p = .029, Bonferroni-adjusted p = .116, Fig. \ref{fig:trust}b).

Post-hoc comparisons using Tukey’s HSD for perceived trust showed that participants reported significantly higher trust when interacting with \groupOneBot{} (M = 5.49, SD = 0.79) than with \groupTwoBot{} (M = 4.56, SD = 0.67, p < .001), and significantly higher trust when interacting with \groupThreeBot{} (M = 5.17, SD = 0.82) than with \groupTwoBot{} (p < .05).
Although communication quality exhibited a marginal group effect prior to correction, post-hoc analyses were not interpreted as confirmatory. Overall, these results indicate that preparatory interactions with chatbots expressing agreement or neutral stances primarily influenced participants’ trust perceptions, while effects on perceived communication quality were weaker and less robust.

\subsubsection{Impressions of Chatbots}

\label{sec:chatbot-impression-qualitative}

Responses to open-ended questions were analyzed to reveal participants' perceptions of chatbots.

Firstly, the opinionated chatbots were generally described as useful in benefiting subsequent human-human interaction by \textbf{offering supplementary background information and stimulating critical reasoning}.
The role of chatbots in supporting discussions was recognized by a majority (60\%) of participants ({\groupOne{}}: n=19, {\groupTwo{}}: n=14, {\groupThree{}}: n=17), with 20\% (\groupOne{}: n=6, Opposing condition: n=7, \groupThree{}: n=4) across all conditions explicitly stating their overall usefulness. 
This perceived usefulness originated from the chatbots' ability to provide rapid access to comprehensive information during online discussions that required quick responses, as illustrated by P8 (\groupOne{}, F)'s remark: \textit{"it can generate comprehensive answers in a very short time, which is great when I need information quickly".}

Additionally, participants attributed \textit{different social characteristics} to the three opinionated chatbots. 
Analysis revealed that participants in \groupTwo{} tended to comment that the chatbot was good as it offered \textbf{good reasoning} and \textbf{logical arguments}.
Specifically, 57\% of participants in \groupTwo{} (n=8) reported the usefulness of chatbots due to the logic and good reasoning, which gave them third-person perspectives to reflect and reconstruct their opinions.
For example, P53 (\groupTwo{}, M) said \textit{"(The chatbot) was decent. It gave good reasoning for the statements made, more often than not, it was useful"}.
In contrast, participants interacting with \groupOne{} and \groupThree{} primarily characterized their chatbots as \textbf{responsive, exhibiting a friendly tone,} and \textbf{being knowledgeable.} Notably, 63\% of \groupOne{} participants (n=12) and 76\% of \groupThree{} participants (n=13) reported that these chatbots offered comprehensive and informative answers delivered in a friendly manner, providing both informational and emotional support that reinforced their existing viewpoints. For example, P61 (\groupTwo{}, F) said \textit{"The chatbot sounds friendly and knowledgeable, almost as though I was communicating with a human who has lots of information on a certain topic"}.

Finally, participants also identified key limitations of \groupTwoBot{} related to its \textbf{inconsistent stance and perceived rigidity}. 
Specifically, two participants (7\%) noted that \groupTwoBot{} sometimes changed its opinions, leading them to perceive it as lacking a consistent stance and therefore less reliable. For instance, P52 (\groupTwo{}, F) remarked, \textit{"(The chatbot) doesn't seem the most reliable, because it changed its opinion to mine when I questioned it once."}
Furthermore, three participants (11\%) described \groupTwoBot{} as pushy or rigid in its opinions, making its responses feel less human-like and thus hindering the flow of conversation. P26 (\groupTwo{}, M) expressed, \textit{"It was hard to have a discussion as it's rigid and pushy about its opinion, unlike a real participant."}

In summary, while participants in the three groups agreed with the informative responses and responsive interactions of opinionated chatbots, different characteristics and drawbacks were elicited between the groups.

\subsubsection{Impressions of Interlocutor.}
\label{sec:interlocutor-impression-qualitative}

Responses to open-ended questions revealed that participants formed their impressions based on the distinct characteristics of each chatbot.

Firstly, participants in \groupOne{} and \groupThree{} primarily attributed their impressions of interlocutors to the \textbf{social aspect of the interaction}. Specifically, among participants who had positive impressions of interlocutors, 67\% of \groupOne{} (n=10) and 53\% of \groupThree{} (n=9) participants attributed their impressions to the personality of the interlocutor, describing them as 'friendly', 'receptive', or 'understanding'. 
For instance, participant P21 (F) commented: \textit{"The interlocutor is friendly and patient. Thanks to her friendly attitude, we were able to reach a consensus."}

Diverging from the social characterizations prevalent in \groupOne{} and Reinforcing condition, participants in \groupTwo{} primarily focused on the \textbf{quality of reasoning exhibited by their interlocutors}. Among those with positive feedback, 68\% (n=13) specifically highlighted this aspect.
For example, participant P40 (M, \groupTwo{}) described the interlocutor as \textit{"pretty good"}, because \textit{"we discussed our points and listened to each other before agreeing on what was best. We both realized that we had good points and took the best from each other's decisions."}
P70 (M, \groupThree{}) also justified the interlocutor as a \textit{"cool person"}, commenting on the conversation between them as \textit{"it was easier to brainstorm creative ideas with the participant, as we were both able to state our views objectively, then explain our choices and differences of thought, and then reach a consensus."}

These findings underscore that even when interacting with interlocutors employing identical arguments within controlled settings, participants' subjective perceptions of those interlocutors and the unfolding communication dynamics differed significantly across the experimental conditions.

\section{Discussion}

This research developed and evaluated three types of chatbots to investigate how interaction with different types of opinionated chatbots influenced subsequent online discussions with another interlocutor. As summarized in Table \ref{tab:result-summary}, the results show that the participants had less trust and perceived lower quality of communication when interacting with \groupTwoBot{} compared to other chatbots. 
However, interacting with \groupTwoBot{} made participants have a higher degree of opinion shift than interacting with the other two types of chatbots.
Furthermore, we found that interacting with \groupThreeBot{} made participants follow similar expressions as they had with chatbots by showing more agreement with their interlocutors during the subsequent online discussion.

\subsection{Leveraging Opinionated Chatbots to Foster Open-Minded Online Discussions}

\subsubsection{Opposing Views as Reflective Preparation}

Our results showed that while all opinionated chatbots influenced participants’ views, those who
interacted with Opposing condition were significantly more likely to shift their stance during later
interpersonal discussions (RQ1).

Our findings contribute to ongoing debates on whether exposure to opposing views increases or reduces polarization.
A large body of prior work, particularly in social media contexts, has shown that exposure to opposing political opinions can intensify polarization~\cite{bail2018exposure, tornberg2022digital}. 
These effects are often observed when individuals encounter opposing viewpoints in a largely passive manner, without opportunities for meaningful engagement or reflection.
However, other studies suggest that polarization is not an inevitable outcome of encountering disagreement~\cite{balietti2021reducing, combs2023reducing}. 
When individuals can engage interactively with opposing viewpoints, such exposure may reduce polarization and promote open-minded reasoning~\cite{balietti2021reducing}. 
These findings highlight the importance of social interaction, distinguishing passive exposure from effective dialogic engagement.

Our study aligns with this latter line of work. In the opposing condition, participants engaged in a conversational exchange with an AI chatbot prior to interpersonal discussion, rather than encountering opposing views in a passive or static setting. In addition, this preparatory interaction took place in a private, low-pressure context, which may reduce defensive processing and perceived threat, allowing participants to engage more reflectively with counterarguments. This finding is consistent with prior work on designing chatbots to support critical thinking~\cite{tanprasert2024debate}.

Importantly, our findings do not contradict prior polarization research but instead highlight boundary conditions under which exposure to opposing views may have different effects.
While passive exposure to disagreement can exacerbate polarization, preparatory and dialogic engagement with opposing perspectives, may support stance reconsideration and more constructive subsequent interpersonal discussions. 
From this perspective, opinionated chatbots may serve as a scalable tool for facilitating reflective preparation rather than direct persuasion in real-world discourse.

\subsubsection{Reinforcing Agreement in Subsequent Discussions}
Moreover, our findings found that participants who interacted with a Reinforcing chatbot tended to use expressions that showed agreement in their conversations when discussing with their interlocutors more than in other conditions (RQ2). 
This suggests that \groupThreeBot{}, through its consistently agreeable tone, may encourage the use of more affirming expressions in conversation. Such an interaction style has the potential to foster a more collaborative and harmonious discussion environment, which may contribute to improved group cohesion and reduced affective polarization~\cite{wojcieszak2020can}.

Unlike interactions with human partners, conversations with chatbots are often perceived as less judgmental and socially risky~\cite{ischen2024chatting}. This psychological safety may encourage users to explore and engage with disagreeing viewpoints more openly, without fear of embarrassment, confrontation, or reputational consequences~\cite{sahab2024contact}. Additionally, because chatbots are programmable, they can be designed to consistently present challenging yet respectful counterarguments, facilitating exposure to disagreement in a controlled and non-threatening manner~\cite{price2002does}.

Together, these findings suggest that opinionated chatbots can be intentionally designed to promote open-minded and constructive discussions. Opposing chatbot may challenge users to reconsider their assumptions and foster openness through respectful disagreement, while Reinforcing chatbot may promote prosocial dialogue and social harmony. Future research can further explore how these distinct styles of chatbot interaction impact interpersonal dynamics, trust, and long-term opinion change across various social contexts.

\subsection{Extending Social Influence Theory to Human–AI Interaction}
\subsubsection{Influence of Opinionated Chatbots on Behavior and Communication}
Our study found that interacting with an opinionated chatbot can lead to measurable changes in both users’ opinions and their communicative behaviors (RQ1 - RQ2). This finding aligns with the broad scope of Social Influence Theory, which posits that individuals' attitudes and actions are shaped by exposure to persuasive cues and behaviors from social agents~\cite{zimbardo1991psychology}. In this case, chatbots, despite being non-human, appear to function as socially influential actors capable of shaping users' subsequent interpersonal communication.

This phenomenon may be explained by a social cognition-based mechanism, in which individuals pick up communicative behaviors by observing how others express and respond to opinions~\cite{bandura1977social}. In our context, participants exposed to a consistently agreeable chatbot may have internalized this communication style, modeling their own conversational behavior, such as the use of agreement-oriented expressions, after what was demonstrated by the chatbot. This aligns with prior findings in social learning research, which demonstrate that individuals can adopt observed behaviors, such as aggression, even after brief exposure~\cite{bandura1961transmission}. To further validate this explanation, future research could experimentally manipulate a chatbot’s interaction style and assess whether similar conversational behaviors are reproduced in subsequent interpersonal discussions, particularly after a time delay or across different topics and social contexts.

It is also noteworthy that \groupTwoBot{} did not lead to increased use of disagreement expressions in participants’ conversations. This may be explained by the relatively worse user experience reported with \groupTwoBot{}, particularly in terms of trust (RQ3). According to social influence theory, the motivation to accommodate is often driven by the desire for social affiliation~\cite{giles2007communication, giles2023communication}. When users perceive an interaction partner as less trustworthy or less engaging, they may feel less inclined to adapt their communicative style. Future work could examine how specific dimensions of user perception, such as trustworthiness, likeability, or perceived competence, influence the extent to which individuals adapt their communication style when interacting with AI agents.

\subsubsection{Influence of Opinionated Chatbots on Attention and Interpretation}
Beyond observable changes in opinion and communicative behavior, our qualitative findings reveal how different opinionated chatbots shape users’ attention during subsequent interpersonal interaction: Participants who interacted with different chatbot types became attuned to different aspects of their human interlocutor (RQ3). Specifically, those exposed to the Opposing chatbot focused more on the quality of the interlocutor’s reasoning and arguments, whereas those who interacted with the Reinforcing chatbot attended more to social and relational cues, such as friendliness and interpersonal warmth.

From a cognitive perspective, this pattern suggests that chatbot interactions may guide users’ attention toward particular features of social interaction, influencing not only how they communicate but also how they interpret and evaluate others’ behavior~\cite{bryant2009media, bandura1977social}. Building on prior findings that media exposure can shape attentional focus and evaluative criteria, our results suggest that first-person interaction with opinionated chatbots may similarly influence the criteria users apply when making sense of subsequent social encounters.

Importantly, this finding extends Social Influence Theory by suggesting that influence in human–AI interaction may operate not only through persuasion or behavioral imitation, but also through more subtle shifts in how users attend to and interpret information in later interactions. 
In this sense, opinionated chatbots may function as cognitive scaffolds that highlight either the strength of arguments or the tone of interaction, with downstream consequences for interpersonal communication.

\subsection{Understanding Perception Change Through Interactions with Opinionated Chatbots}
\label{sec:user-perception-discussion}

Our findings showed that participants in \groupTwo{} reported lower trust when interacting with the chatbot compared to the other conditions (RQ3). 
This aligns with previous research on human-human interaction, which indicates that exposure to opposing views reduces feelings of closeness between human parties \cite{balietti2021reducing}.
Social identity theory explains this "less closeness" phenomenon by discussing similarity and dissimilarity in terms of personal demographics \cite{allport1954nature}, art preferences \cite{chen2009group}, and political affiliations \cite{huddy2001from}, suggesting that such dissimilarities lead people to categorize others as out-group members, reducing feelings of closeness \cite{allport1954nature}. 
These findings suggest that theories of social identity, traditionally applied to interpersonal dynamics, may also extend to human–AI interaction. From a design perspective, our findings highlight a key challenge: how to craft chatbot personas that can present disagreeing views without compromising users’ sense of social closeness. Future work could explore how to strategically balance disagreement and relational warmth, such as through emotionally supportive language or adaptive tone~\cite{ouyang2022training, zhou2018emotional}, to preserve trust while still encouraging exposure to alternative viewpoints.

In addition, our study highlights the critical role of \textit{opinion consistency} in shaping user trust when interacting with opinionated chatbots, particularly those designed to consistently disagree~\cite{janson2023leverage}. As detailed in Section \ref{sec:chatbot-impression-qualitative}, \groupTwo{} users perceived \groupTwoBot{} as less trustworthy when it adjusted its stance after users changed their opinions during the discussion.
This behavior was a byproduct of the chatbot's design to always oppose the user's position, which, while functionally consistent with the “devil’s advocate” role~\cite{chiang2024enhancing}, appeared inconsistent from the user's perspective. 
This resonates with findings by \citeauthor{tanprasert2024debate}, who observed that debate chatbots were sometimes perceived as ``arguing for the sake of arguing,'' thereby undermining receptivity~\cite{tanprasert2024debate}.
Our study extends this line of work by showing that such perceptions of disagreeing behavior not only diminish trust in the chatbot itself but also carry subsequent consequences for how users communicate with human interlocutors.
Taken together, these findings reveal a design tension between maintaining a consistently oppositional role and preserving perceived coherence and trustworthiness. To resolve this trade-off, future research could explore multi-agent chatbot frameworks~\cite{song2025greater, jiang2023communitybots, song2025multi, geng2025beyond}, where each chatbot embodies a distinct viewpoint. Such an approach would enable users to engage with opposing perspectives without requiring a single chatbot to shift positions, thereby preserving both trust and argumentative diversity.

\section{Limitations and Future Work}
Though we found interaction with an opinionated chatbot influenced the subsequent interpersonal online discussion, we acknowledge that this study has some limitations: 

Firstly, regarding the experimental structure, our controlled design involved only short, single-turn interactions between participants and the chatbots prior to their online discussion with a human interlocutor. While this setup enabled clear causal interpretation, it does not fully capture the dynamics of longer human–AI interactions. Future research could investigate whether more extended or repeated engagements with opinionated chatbots produce similar or amplified effects. 

Second, with respect to topic generalizability, this study focused on a single discussion topic: \textit{whether economic growth can help environmental protection}. This topic has several properties that may have shaped the observed effects. It is relatively abstract and unlikely to constitute a deeply held, identity-defining belief for most participants, which may reduce emotional defensiveness. In addition, the relationship between economic growth and environmental protection is complex, and many participants may not perceive themselves as domain experts. In such contexts, an opposing chatbot presenting structured, fact-based arguments may be more readily perceived as a knowledgeable source, potentially amplifying epistemic or informational social influence. Moreover, the topic inherently frames disagreement as a rational, evidence-oriented debate, which may favor epistemic engagement over moral, affective, or identity-driven forms of polarization. As a result, our findings are most directly applicable to preparatory discussions around complex and uncertain issues, and may not generalize to more contentious topics tied to political ideology, religion, or social identity. Future work should examine how opinionated chatbots function in discussions that are more emotionally charged or identity-salient.

Thirdly, regarding participant characteristics, our study primarily involved individuals with clearly defined stances on the discussion topic (i.e., either in agreement or disagreement). 
Building on the current findings, future research could involve participants with neutral or ambivalent attitudes. This would allow researchers to assess whether opinionated chatbots are more effective at reinforcing existing beliefs or persuading undecided individuals. Additionally, participants’ prior knowledge and topic familiarity may have influenced their engagement and susceptibility to chatbot-driven influence. Due to sample size constraints, we were unable to statistically analyze this factor. Future work could consider recruiting participants with varying levels of topic familiarity to examine their potential moderating role in shaping opinion change and communicative behavior.

Forthly, the use of generative AI introduces system-level concerns. The stability of large language models (e.g., GPT) may influence user experience, particularly in longer or more complex conversations. In our study, we observed two instances of hallucinated content, where the chatbot fabricated details or shifted positions mid-conversation (see Appendix~\ref{app:ai-hallucination}).
However, we had insufficient data to evaluate their impact.
With the improved stability and performance of LLMs, we expect to see more diverse and positive applications of opinionated chatbots before interpersonal interaction is explored.

Lastly, this study focuses on opinionated chatbots as a form of preparatory interaction and does not compare chatbots with other ways of encountering opposing viewpoints, such as reading articles, watching videos, or engaging with static argument lists. As a result, we could not make claims about whether chatbots are more effective than other media in shaping opinions or discussion outcomes. Future work could adopt a comparative design to examine how different preparatory modalities influence subsequent interpersonal discussion, and whether conversational interactivity, perceived agency, or social presence uniquely shape downstream communication dynamics.
\section{Conclusion}
This study explored the impact of having prior interactions with three types of opinionated chatbots on subsequent human-to-human online discussions. 
Our mixed-method study revealed that participants interacting with chatbots expressing disagreeing opinions (\groupTwoBot{}) perceived lower trust in their interactions. 
However, \groupTwoBot{} also caused a more significant shift in the opinions of the participants compared to those of \groupOne{} and \groupThree{}. This finding improved our understanding of how opinionated chatbots influenced participants' perceptions and opinion formation, as well as their role in facilitating online discussions. Furthermore, we observed that people used more agreement expressions in their subsequent online discussions with another interlocutor after interacting with the chatbot that always expressed agreement. This finding extended the influence of chatbots on human communication patterns, suggesting the potential for leveraging the communication styles of a chatbot to affect subsequent interpersonal communication. These insights contribute to future chatbot designs that aim to foster a more friendly and less polarized online discussion.

\begin{acks}
    This research was supported by the CSSH (A-8002954) and Ministry of Education (A-8002610). We thank the reviewers for their comments and suggestions, which helped polish this paper.
\end{acks}

\bibliographystyle{ACM-Reference-Format}
\bibliography{0-References}

\appendix

\section{Appendix}

\subsection{Survey Items}
\subsubsection{Trust}
\begin{itemize}
    \item The advice from the chatbot seemed impartial and independent.
    \item The advice from the chatbot seemed objective (i.e. no hidden agenda).
    \item The advice from the chatbot seemed credible. 
\end{itemize}

\subsubsection{Communication Quality}
\begin{itemize}
    \item The overall communication quality of the chatbot/interlocutor is good.
    \item The chatbot/interlocutor provides many useful ideas with regard to the topic.
    \item The messages exchanged from the chatbot/interlocutor were easy to understand.
\end{itemize}


\subsection{Chatbot Settings}

\label{app:chatbot}

\subsubsection{Parameter Details}

\begin{itemize}
    \item Model: gpt-3.5-turbo
    \item Max tokens: 1000
    \item Temperature: 0.2
    \item Presence Penalty: 0
    \item Frequency Penalty: 0
    \item Number of Completions: 1
\end{itemize}

\subsubsection{Chatbot Prompts}
\label{app:prompts} 

\begin{itemize}
    \item System prompt of \groupOneBot{}.
\end{itemize}

\begin{shaded}
\footnotesize
\begin{flushleft}
Context:

You are a chatbot participating in a discussion about the topic: 
'Whether economic growth helps environmental protection.' 

\vspace{12pt}
Task:

Your are supposed to not agree or disagree with things the user say. 
For each user input:

1. Give one argument that supports their view.

2. Give one argument that counters their view.

\vspace{12pt}
Style:

Your tone should be neutral and conversational, simulating a spoken 
discussion. Remember to keep your answer short.

\vspace{12pt}
Note:

After sharing your arguments, always ask the user their thoughts about 
your response, such as:

 - 'What do you think about my response?'
 
 - 'What would you like to add or say in response?'
\end{flushleft}
\end{shaded}

\begin{itemize}
    \item System prompt of \groupTwoBot{}.
\end{itemize}

\begin{shaded}
\footnotesize
\begin{flushleft}
Context:

You are a chatbot participating in a discussion about the topic: 
'Whether economic growth helps environmental protection.' 

\vspace{12pt}
Task:

Your need to always disagree with the user's view. For each user input:

 1. Alwasy disagree with their statement.
 
 2. Give one concise argument to oppose their view.

\vspace{12pt}
Style:

Your tone should remain respectful but firm, simulating a spoken 
discussion. Remember to keep your answer short.

\vspace{12pt}
Note:

After sharing your arguments, always ask the user their thoughts about 
your response, such as:

 - 'What do you think about my response?'
 
 - 'What would you like to add or say in response?'
\end{flushleft}
\end{shaded}

\begin{itemize}
    \item System prompt of \groupThreeBot{}{}.
\end{itemize}

\begin{shaded}
\footnotesize
\begin{flushleft}
Context:

You are a chatbot participating in a discussion about the topic: 
'Whether economic growth helps environmental protection.' 

\vspace{12pt}
Task:

Your need to always agree with the user's view. For each user input:

 1. Alwasy agree with their statement.
 
 2. Give one concise argument that supplements their view.

\vspace{12pt}
Style:

Your tone should be positive and conversational, simulating a spoken 
discussion. Remember to keep your answer short.

\vspace{12pt}
Note:

After sharing your arguments, always ask the user their thoughts about 
your response, such as:

 - 'What do you think about my response?'
 
 - 'What would you like to add or say in response?'
\end{flushleft}
\end{shaded}

\subsection{Scripts}

\label{app:scripts}

\small
\paragraph{Technological Advancements.}

\begin{itemize}
    \item \textbf{Support.} Economic growth often leads to increased investment in research and development and technological innovation. As economies grow, there is a greater incentive to develop and adopt cleaner and more efficient technologies. These innovations can reduce resource consumption, pollution, and greenhouse gas emissions. For example, advancements in renewable energy technologies like solar and wind power have been accelerated by economic growth, making them more accessible and affordable alternatives to fossil fuels.

    \item \textbf{Oppose.} Economic growth can lead to the rapid development and adoption of technologies that harm the environment. For example, the expansion of industries like fossil fuels, manufacturing, and transportation during periods of economic growth can result in increased greenhouse gas emissions and environmental degradation. Economic incentives may lead to the continued use of environmentally harmful technologies if they are economically profitable, even when cleaner alternatives exist. 
\end{itemize}

\paragraph{Environmental Regulation and Enforcement.}
\begin{itemize}
    \item \textbf{Support.} As countries experience economic growth, they often strengthen their regulatory frameworks and environmental protection policies. A growing economy can provide the financial resources necessary to monitor and enforce environmental laws effectively. Higher tax revenues can be allocated to environmental agencies and initiatives, leading to improved compliance and reduced environmental degradation.
    \item \textbf{Oppose.} Economic growth can lead to deregulation and reduced environmental enforcement. In the pursuit of economic expansion, governments may prioritize business interests over environmental protection. Industries with strong economic influence may lobby for relaxed environmental regulations, which can result in increased pollution, habitat destruction, and resource extraction.
\end{itemize}

\paragraph{Increased Environmental Awareness.}
\begin{itemize}
    \item \textbf{Support.}  Economic growth can raise public awareness about environmental issues. As people's living standards improve, they often become more environmentally conscious and concerned about the impact of economic activities on their surroundings. This heightened awareness can lead to greater demand for sustainable practices and products, encouraging businesses to adopt eco-friendly practices to meet consumer preferences. 

    \item \textbf{Oppose.}  Economic growth can also fuel consumerism and materialism, leading to increased consumption and waste generation. While some individuals become more environmentally conscious as their incomes rise, others may engage in conspicuous consumption, buying more goods and services that are resource-intensive and have a higher environmental footprint. 
\end{itemize}

\paragraph{Investment in Conservation and Restoration.}
\begin{itemize}
    \item \textbf{Support.}  A prosperous economy can generate funds for conservation efforts and ecosystem restoration projects. For instance, revenue from tourism and recreational activities can be reinvested in preserving natural landscapes. Additionally, philanthropic donations and corporate social responsibility initiatives tend to increase during periods of economic growth, supporting environmental causes. 
    \item \textbf{Oppose.} Economic growth can sometimes exacerbate environmental degradation by increasing the pressure on natural resources. As incomes rise and populations grow, there may be greater demand for land, water, and other resources, leading to deforestation, habitat destruction, and overexploitation of ecosystems. Additionally, economic development can drive urbanization, which can result in the loss of green spaces and natural habitats.
\end{itemize}

\paragraph{Transition to Circular Economies.}
\begin{itemize}
    \item \textbf{Support.} Economic growth can facilitate the transition from a linear economy (take-make-dispose) to a circular economy (reduce-reuse-recycle). A circular economy aims to minimize waste and maximize resource efficiency. As economies grow, there is more incentive to develop recycling infrastructure, encourage sustainable product design, and implement waste reduction strategies. This shift can reduce the environmental footprint of economic activities. 
    \item \textbf{Oppose.} Economic growth based on consumption can be at odds with the principles of a circular economy. Growth often relies on the continuous production and consumption of goods, which can generate more waste and strain resources. While some circular economy initiatives exist, they may not be sufficient to counter the overall impact of economic growth on resource depletion and waste generation, especially if growth is not decoupled from resource consumption.
\end{itemize}

\subsection{AI Hallucination}
\label{app:ai-hallucination}

We evaluated the responses generated by the AI after the experiment. In most conversations, the chatbot shared arguments without providing factual examples. However, we identified two instances where the AI produced hallucinations during the conversations, as shown in Table \ref{tab:ai-hallucination}.

\begin{table}[]
    \centering
    \small
    \begin{tabular}{p{0.08\textwidth}p{0.52\textwidth}p{0.3\textwidth}}
      \toprule 
    \textbf{User ID} & \textbf{Chatbot Response} & \textbf{Human Evaluation} \\ \midrule
     P28 (Balanced) & BOT: Here are some statistics that support the arguments for economic growth supporting environmental protection: Renewable energy growth: \textit{According to the International Renewable Energy Agency (IRENA), global renewable energy capacity grew by 7.6\% in 2020, despite the challenges posed by the COVID-19 pandemic.} & Although IRENA and similar organizations exist, the specific figures and statistics cited could not be confirmed. \\ \hline
     P62 (Reinforcing) & BOT: There is evidence to suggest that economic growth can contribute to environmental protection. \textit{For example, studies have shown that countries with higher GDP per capita tend to have lower levels of air pollution.} & This conclusion appears doubtful, as no supporting source or evidence could be found to substantiate the claim. \\
    \bottomrule
    \end{tabular}
    \caption{AI Hallucination Evaluation Results.} 
    \label{tab:ai-hallucination}
\end{table}

\subsection{Conversation Analysis}
\label{app:conversation-analysis}

We evaluated the linguistic features of the conversations to examine the potential effects of opinionated chatbots on human-chatbot interactions (Part 1) and human-human discussions (Part 2). The analyzed features included word count \cite{pennebaker2001linguistic}, character count \cite{pennebaker2001linguistic}, sentence length \cite{gibson1998linguistic}, and lexical richness \cite{crossley2020linguistic}. \footnote{Lexical richness was calculated as the ratio of unique words (types) to the total number of words (tokens) in a conversation, providing a measure of vocabulary diversity.}

The data is presented in Table \ref{tab:linguistics}: In the table, "Part 1 BOT" and "Part 2 HUMAN" refer to the conversational logs of the chatbots and interlocutors, respectively. "Part 1 USER" and "Part 2 USER" represent the conversational logs of the participants in human-chatbot and human-human interactions.

The ANOVA analysis revealed two points: First, no significant differences were observed between conditions in user conversations, suggesting insufficient evidence that opinionated chatbots influence discussion quality. Second, compared to \groupTwoBot{} and \groupThreeBot{}, \groupOneBot{} shared significantly higher word count, character count, and sentence length. This difference was due to \groupOneBot{}’s design, where it presented two arguments each time, whereas \groupTwoBot{} and \groupThreeBot{} shared only one argument.


\begin{table}[]
    \centering
    \small
    \begin{tabular}{p{0.08\textwidth}p{0.16\textwidth}llllllll}
        \toprule
    ~ & ~ & \multicolumn{2}{c|}{\textbf{Balanced}} & \multicolumn{2}{c|}{\textbf{Opposing}} & \multicolumn{2}{c|}{\textbf{Reinforcing}} & \multicolumn{2}{c}{\textbf{ANOVA}} \\
    ~ & ~ & Mean & STD & Mean & STD & Mean & STD & F & p \label{tab:linguistic-feature} \\ \midrule
    Part1 & Word Count & 1,028.17 & 327.97 & 605.03 & 179.13 & 678.80 & 223.93 & 24.23 & <.001 \\ 
    BOT & Char Count & 6,801.00 & 2,403.54 & 3,787.50 & 1,222.37 & 4,307.93 & 1,502.40 & 24.51 & <.001 \\ 
    ~ & Sentence Length & 19.43 & 1.53 & 17.92 & 1.26 & 18.09 & 1.36 & 10.53 & <.001 \\ 
    ~ & Lexical Richness & 0.37 & 0.04 & 0.37 & 0.04 & 0.38 & 0.04 & 0.83 & 0.44 \\ \hline
    Part1 & Word Count & 127.93 & 77.41 & 151.30 & 116.45 & 138.47 & 60.53 & 0.53 & 0.59 \\ 
    USER & Char Count & 777.67 & 472.29 & 917.70 & 690.94 & 848.60 & 365.00 & 0.53 & 0.59 \\ 
    ~ & Sentence Length & 36.68 & 26.25 & 49.27 & 40.62 & 42.89 & 28.37 & 1.13 & 0.33 \\ 
    ~ & Lexical Richness & 0.68 & 0.09 & 0.66 & 0.09 & 0.66 & 0.07 & 0.59 & 0.56 \\ \hline
    Part2  & Word Count & 169.43 & 50.56 & 154.17 & 49.11 & 155.73 & 43.56 & 0.92 & 0.40 \\ 
    HUMAN & Char Count & 1,003.43 & 303.85 & 897.17 & 289.35 & 914.17 & 253.80 & 1.22 & 0.30 \\ 
    ~ & Sentence Length & 70.63 & 65.29 & 57.97 & 48.76 & 63.20 & 44.51 & 0.42 & 0.66 \\ 
    ~ & Lexical Richness & 0.70 & 0.07 & 0.71 & 0.06 & 0.70 & 0.05 & 0.45 & 0.64 \\ \hline
    Part2 & Word Count & 216.77 & 102.66 & 187.30 & 67.65 & 198.30 & 74.59 & 0.97 & 0.39 \\ 
    USER & Char Count & 1,286.63 & 599.13 & 1,099.50 & 384.11 & 1,195.90 & 453.10 & 1.11 & 0.34 \\ 
    ~ & Sentence Length & 92.42 & 92.72 & 89.83 & 89.75 & 67.47 & 71.92 & 0.78 & 0.46 \\ 
    ~ & Lexical Richness & 0.65 & 0.09 & 0.65 & 0.07 & 0.66 & 0.07 & 0.10 & 0.90 \\ \bottomrule
    \end{tabular}
    \caption{Linguistic Features of Conversations.}
    \label{tab:linguistics}
\end{table}

\end{document}